\documentclass[12pt]{article}
\pdfoutput=1
\usepackage{amsmath}
\usepackage{color}
\def\a{\alpha}
\def\b{\beta}

\def\d{\delta}

\def\g{\gamma}

\def\be{\begin{equation}}
\def\ee{\end{equation}}
\def\arr{\begin{array}{rll}}
\def\ea{\end{array}}
\def\bea{\begin{eqnarray}}
\def\eea{\end{eqnarray}}
\def\N2{$N{=}2$}

\def\>{\rangle}
\def\<{\langle}
\def\+{\dagger}
\def\={\ =\ }

\begin{document}
%\large
\renewcommand{\thefootnote}{\fnsymbol{footnote}}
\begin{titlepage}
\setcounter{page}{0}

\begin{center}
{\LARGE\bf $\mathcal{N}=1,2,3$ $\ell$--conformal Galilei superalgebras }\\

\vskip 1.5cm
$
\textrm{\Large Anton Galajinsky}^{a}  ~  \textrm{\Large  and Ivan Masterov}^{b}
$
\vskip 0.7cm
${}^{a}${\it
Tomsk State University of Control Systems and Radioelectronics, 634050 Tomsk, Russia} \\
\vskip 0.2cm
${}^{b}${\it
Tomsk Polytechnic University, 634050 Tomsk, Lenin Ave. 30, Russia} \\
\vskip 0.2cm
{e-mails:
a.galajinsky@tusur.ru, masterov@tpu.ru}
\vskip 0.5cm

\end{center}
\vskip 1cm
\begin{abstract} \noindent
The issue of constructing $\mathcal{N}=1,2,3$ supersymmetric extensions of the $\ell$--conformal Galilei algebra is reconsidered following the approach in [JHEP 1709 (2017) 131]. Drawing a parallel between
acceleration generators entering the superalgebra and irreducible supermultiplets of $d=1$, $\mathcal{N}$--extended superconformal group,
a new $\mathcal{N}=1$ $\ell$--conformal Galilei superalgebra, two new $\mathcal{N}=2$ variants, and two new $\mathcal{N}=3$ versions are built. Realisations in terms of differential operators in superspace are given.

\end{abstract}


\vskip 0.5cm


\end{titlepage}

\renewcommand{\thefootnote}{\arabic{footnote}}
\setcounter{footnote}0

\noindent
{\bf 1. Introduction}\\

\noindent
The exploration of the non--relativistic version of the AdS/CFT--correspondence initiated in \cite{DTS,BM} generated a great deal of interest in non--relativistic conformal (super)groups and field theories enjoying such symmetries (earlier developments are reviewed in \cite{MT}, for more recent applications see \cite{CHOR,HKO,HHMOY} and references therein). In particular, the holographic dictionary was extended to encompass strongly coupled condensed matter systems (see \cite{NS} and references therein).

A peculiar feature of non--relativistic conformal transformations is that temporal and spatial coordinates scale differently under dilatations. If one is concerned with maintaining the full Galilei algebra, the most general {\it finite--dimensional} conformal extension is given by the so called $\ell$--conformal Galilei algebra \cite{Henkel,NOR}
\be\label{CGA}
[L_m,L_n]=-{\rm i} (m-n) L_{m+n}, \qquad
[L_n,{\bf U}_p]=-{\rm i} \left( \ell n-p\right) {\bf U}_{n+p},
\ee
where $n=-1,0,1$, $p=-\ell,\ldots,\ell$ and $\ell$ is a non--negative (half)integer parameter. $L_{-1}$, $L_0$, $L_1$ generate time translations, dilatations, and special conformal transformations, respectively, which all together form $sl(2,R)$ subalgebra. ${\bf U}_{-\ell}$ describes spatial translations, ${\bf U}_{-\ell+1}$ is linked to Galilei boosts, while higher values of $p$ are commonly associated with constant accelerations.\footnote{In what follows, we call ${\bf U}_p$ the acceleration generators for short. $so(d)$ subalgebra
$[M_{\a\b},M_{\g\d}]={\rm i}(\d_{\a\g} M_{\b\d}+\d_{\b\d} M_{\a\g}-
\d_{\b\g} M_{\a\d}-\d_{\a\d} M_{\b\g})$ entering the Galilei algebra was omitted in (\ref{CGA}). Throughout the paper, generators carrying vector indices with respect to $so(d)$ appear in boldfaced type, e.g. ${\bf A}=A_\alpha$, $\alpha=1,\dots,d$, which obey $[M_{\a\b},A_\g]={\rm i}(\d_{\a\g} A_\b-\d_{\b\g} A_\a)$.}  In accord with (\ref{CGA}), $L_n$ and ${\bf U}_p$ have conformal weights $1$ and $\ell$, respectively.

In modern literature, the reciprocal of $\ell$ is called the rational dynamical exponent and (\ref{CGA}) is sometimes referred to as the conformal Galilei algebra with rational dynamical exponent. The instances $\ell=\frac 12$ and $\ell=1$, known as the Schrodinger algebra and the conformal Galilei algebra, have received the utmost attention (for a review see \cite{DH1}).

If the tower of acceleration generators is reduced to a single spatial translation generator, the special conformal transformations are discarded, and the dynamical exponent is regarded arbitrary, one recovers the so called Lifshitz algebra. The Lifshitz holography has interesting peculiarities and it has been extensively studied in the past (see \cite{CHOR,HKO,KLM,MT1} and references therein).

When constructing a specific
dynamical realisation of a non--relativistic conformal group, generators of the corresponding Lie algebra are linked to constants of the motion. Because the number of functionally independent integrals of motion needed to integrate a differential equation correlates with its order, dynamical
realisations of an $\ell> \frac 12 $ conformal Galilei algebra in mechanics in general involve higher derivative terms. In particular, symmetries of a higher order free particle were studied in \cite{GKa} while \cite{AGGM} established the $\ell$--conformal Galilei symmetry of the Pais--Uhlenbeck oscillator. Field theories with $\ell>\frac 12$ conformal Galilei symmetry remain almost completely unexplored (see however the recent studies in \cite{HLO,CS}).

Turning to supersymmetric extensions of the $\ell$--conformal Galilei algebra, there are several competing approaches to be mentioned. One can either consider a relativistic superconformal algebra in a suitable dimension and analyse its subalgebras, or implement the non--relativistic contraction, or extend $d=1$, $\mathcal{N}$--extended superconformal algebra by acceleration generators in a proper way (see e.g. \cite{DH,HU,SY,AL,SS,FL} for $\ell=\frac 12,1$ and \cite{IM,A,AKT,GM,GK} for an arbitrary value of $\ell$).

Given a supersymmetric generalisation of (\ref{CGA}), $sl(2,R)$ subalgebra formed by $L_n$ is enlarged to a proper $d=1$, $\mathcal{N}$--extended superconformal algebra, while the acceleration generators ${\bf U}_p$ are accompanied by extra superpartners. In general, the acceleration generators and their superpartners combine together to form an irreducible supermultiplet of the $d=1$, $\mathcal{N}$--extended superconformal algebra. It is customary to denote such supermultiplets by the triplet $(\mathcal{N}-\mathcal{A},\mathcal{N},\mathcal{A})$, in which $\mathcal{N}-\mathcal{A}$ designates the number of physical real bosons, $\mathcal{N}$ counts the number of physical real fermions, and $\mathcal{A}$ quantifies auxiliary (non--propagating) real bosons. The full number of bosons equals that of fermions.  In particular, the $\mathcal{N}=1$ $\ell$--conformal Galilei superalgebra in \cite{A} introduces one fermionic partner of ${\bf U}_p$, which together form a $(1,1,0)$ supermultiplet. The $\mathcal{N}=2$ superalgebras in \cite{IM,A,AKT} link to $(1,2,1)$ and $(2,2,0)$ supermultiplets. $\mathcal{N}=4$ superalgebras associated with various $d=1$, $\mathcal{N}=4$ supermultiplets, including the most general case of the exceptional supergroup $D(2,1;\alpha)$, were investigated in \cite{GM,GK}. To the best of our knowledge, the $\mathcal{N}=3$ case has not yet been explored for an arbitrary value of the parameter $\ell$.

A particularly efficient method of building an $\mathcal{N}$--extended $\ell$--conformal Galilei superalgebra, which encompasses the previous studies in \cite{IM,A,AKT,GM}, was proposed in \cite{GK}. Given an $(\mathcal{N}-\mathcal{A},\mathcal{N},\mathcal{A})$ supermultiplet of the $d=1$, $\mathcal{N}$--extended superconformal algebra, one identifies the first member of the triplet with the bosonic acceleration generator ${\bf U}$. In general, it carries an extra index which controls the number $\mathcal{N}-\mathcal{A}$ chosen. Computing the commutator of ${\bf U}$ with the supersymmetry generators entering the $d=1$, $\mathcal{N}$--extended superconformal algebra, one obtains $\mathcal{N}$ superpartners of ${\bf U}$, say ${\bf S}$. The anticommutator of ${\bf S}$ with the supersymmetry generators then yields a combination of a new auxiliary bosonic acceleration generator, say ${\bf A}$, and the original ${\bf U}$. Finally, the commutator of ${\bf A}$ and the supersymmetry generators may only result in ${\bf S}$. The number coefficients accompanying the generators on the right hand side of the structure relations are fixed by demanding that the ensuing superalgebra is finite--dimensional and that the super Jacobi identities are satisfied.

Apart from the indices which keep the balance between the components of an $(\mathcal{N}-\mathcal{A},\mathcal{N},\mathcal{A})$ supermultiplet, the acceleration generators carry extra indices specifying the $sl(2,R)$ transformation rules. In general, given the triplet $({\bf U},{\bf S},{\bf A})$, one can either assign to its members conformal weights $(\ell,\ell+\frac 12,\ell+1)$,
or alternatively choose the descending sequence $(\ell,\ell-\frac 12,\ell-1)$ \cite{GK}. Thus, for each value of the parameter $\ell$ one can construct two variants of an $\mathcal{N}$--extended $\ell$--conformal Galilei superalgebra, which in general are not isomorphic.

The goal of this work is to reconsider the issue of constructing $\mathcal{N}=1,2,3$ $\ell$--conformal Galilei superalgebras following the method in \cite{GK}.

In the next Section, two $\mathcal{N}=1$ $\ell$--conformal Galilei superalgebras are built by enlarging the $osp(1|2)$ superconformal algebra with a bosonic acceleration generator of conformal weight $\ell$ and its superpartner which either has conformal weight $\ell-\frac 12$ or $\ell+\frac 12$. The former case corresponds to the superalgebra in \cite{A}, while the latter is new.

Sect. 3 is devoted to $\mathcal{N}=2$ $\ell$--conformal Galilei superalgebras which build upon $su(1,1|1)$. Variants associated with $(1,2,1)$ and $(2,2,0)$ irreducible supermultiplets are considered. In the former case, a new $\mathcal{N}=2$ superalgebra is obtained which, in addition to $\ell$, involves an arbitrary real parameter $a$. The latter measures the $U(1)$ charge of the bosonic acceleration generators. It is shown that the $\mathcal{N}=2$ superalgebra introduced in \cite{IM} corresponds to setting the parameter $a$ to zero and choosing the descending sequence of conformal weights for the acceleration generators. For a $(2,2,0)$ supermultiplet, we reproduce the superalgebra in \cite{AKT} and construct its twin copy.

In Sect. 4, $\mathcal{N}=3$ $\ell$--conformal Galilei superalgebras are studied. It is argued that the variants associated with irreducible supermultiplets of the type $(1,3,2)$ and $(3,3,0)$ (other options follow by redefinition of $\ell$) are incompatible with the super Jacobi identities. Yet, introducing extra acceleration generators, such that the full set fits a component content of a real bosonic scalar superfield in $\mathcal{R}^{1|3}$ superspace, one can circumvent the problem. Two new $\mathcal{N}=3$ $\ell$--conformal Galilei superalgebras of such a kind are built, which link to a reducible supermultiplet of $OSp(3|2)$.

In Appendix, realisations of the $\mathcal{N}=1,2,3$ $\ell$--conformal Galilei superalgebras in terms of differential operators in superspace are given.

Throughout the paper, summation over repeated indices is understood.

\vspace{0.5cm}

\noindent
{\bf 2. $\mathcal{N}=1$ $\ell$--conformal Galilei superalgebras}\\

\noindent
Before we proceed to the construction of the $\mathcal{N}=1$ $\ell$--conformal Galilei superalgebras,
let us fix the structure relations of $osp(1|2)$. They are most easily obtained by focusing on the Hamiltonian formulation of a free superparticle in $\mathcal{R}^{1|1}$ superspace. Introducing a real bosonic canonical pair $(x,p)$ and a self--conjugate real fermion $\theta$, which obey the Poisson brackets $\{x,p\}=1$, $\{\theta,\theta \}=-{\rm i}$, one can readily verify that the functions
\be\label{osp12}
L_{-1}=\frac{1}{2} p^2, \qquad L_0=-\frac 12 xp, \qquad L_1=\frac 12 x^2, \qquad
Q_{-\frac 12}=-p \theta, \qquad Q_{\frac 12}=x \theta,
\ee
form a closed superalgebra. Promoting the Poisson brackets to (anti)commutators, one arrives at the structure relation of $osp(1|2)$
\be\label{alg0}
[L_m,L_n]=-{\rm i} (m-n) L_{m+n}, \qquad [L_n,Q_r]=-{\rm i} \left(\frac{n}{2}-r \right) Q_{n+r}, \qquad
\{Q_r,Q_s \}=2 L_{r+s},
\ee
where $L_n$, with $n=-1,0,1$, form $sl(2,R)$ subalgebra and $Q_r$, with $r=-\frac 12, \frac 12$, encode the supersymmetry generator and its superconformal partner. $(L_n,Q_r)$ are assumed to be Hermitian operators. In accord with (\ref{alg0}), they have conformal weights $1$ and $\frac 12$, respectively.

In order to construct an $\mathcal{N}=1$ supersymmetric extension of the $\ell$--conformal Galilei superalgebra, one extends $osp(1|2)$ with the bosonic acceleration generator ${\bf U}_m$ introduced in Eq. (\ref{CGA}) above. Because the commutator of ${\bf U}_m$ and $Q_r$ is Grassmann--odd, it should be regarded as a new fermionic generator, say ${\bf S}_p$. In general, the range of values of the lower index $p$ correlates with conformal weight of ${\bf S}_p$. Two options are available \cite{GK}. One can either assign the conformal weight $\ell+\frac 12$ to ${\bf S}_p$, or alternatively choose $\ell-\frac 12$.  In the former case, one has
\be\label{QU0}
[Q_r,{\bf U}_m]={\rm i} {\bf S}_{r+m}, \qquad [L_n,{\bf S}_p]=-{\rm i} \left( \left( \ell+\frac 12 \right)n-p \right) {\bf S}_{n+p},
\ee
with $p= -\ell-\frac{1}{2},\dots,\ell+\frac{1}{2}$.
The anticommutator of $Q_r$ and ${\bf S}_p$ is then uniquely determined by the range of values of the indices and the super Jacobi identities
\be\label{QS)}
\{Q_r,{\bf S}_p \}=-\left( \left( 2\ell+1 \right)r-p \right) {\bf U}_{r+p}.
\ee
It is assumed that ${\bf U}_m$ and ${\bf S}_p$ (anti)commute with each other.
To the best of our knowledge, this $\mathcal{N}=1$ $\ell$--conformal Galilei superalgebra is new.

A twin copy of the superalgebra arises if one assigns ${\bf S}_p$ with the conformal weight $\ell-\frac 12$. Three structure relations are modified accordingly
\be
[Q_r,{\bf U}_m]={\rm i} (2 \ell r-m) {\bf S}_{r+m}, \quad [L_n,{\bf S}_p]=-{\rm i} \left( \left( \ell-\frac 12 \right)n-p \right) {\bf S}_{n+p}, \quad
\{Q_r,{\bf S}_p \}=- {\bf U}_{r+p},
\nonumber
\ee
with ${\bf S}_p : p= -\ell+\frac{1}{2},\dots,\ell-\frac{1}{2}$, while the rest remains unchanged. Written in different notations, this variant was proposed in \cite{A}.

For reader's convenience, realisations of the $\mathcal{N}=1$ $\ell$--conformal Galilei superalgebras in terms of differential operators in superspace are given in Appendix.

Note that the two versions above are not entirely identical. In particular, the second option involves two less fermionic generators ${\bf S}_p$.
Curiously enough, the latter follows from the former if one implements the formal substitution
\be\label{changeN1}
\ell \to \ell-\frac 12, \qquad
{\bf U}_m \to -{\rm i} {\bf S}_m, \qquad {\bf S}_p \to {\bf U}_p,
\ee
and interchanges commutators with anticommutators where appropriate. As the Grassmann parities of the generators get altered, the change should be regarded as a kind of duality transformation.

Worth mentioning also is that, discarding the lower indices attributed to $sl(2,R)$, the set of the acceleration generators $({\bf U},{\bf S})$ fits a component content of a real scalar superfield $\Phi(x,\theta)=U(x)+ \theta S(x)$ in $\mathcal{R}^{1|1}$ superspace. The latter links to a $(1,1,0)$ supermultiplet of the $OSp(1|2)$ superconformal group.
\vspace{0.5cm}

\noindent
{\bf 3. $\mathcal{N}=2$ $\ell$--conformal Galilei superalgebras}\\

\noindent
{\it 3.1. Structure relations of $su(1,1|1)$}\\

\noindent
An $\mathcal{N}=2$  $\ell$--conformal Galilei superalgebra builds upon $su(1,1|1)$. In order to establish structure relations of the latter, it suffices to consider the Hamiltonian formulation of a free superparticle in $\mathcal{R}^{1|2}$ superspace
\begin{align}\label{N2f}
&
L_{-1}=\frac{1}{2} p^2, && L_0=-\frac 12 xp, && L_1=\frac 12 x^2, &&
Q_{-\frac 12}=-p \theta,
\nonumber\\[2pt]
&
{\bar Q}_{-\frac 12}=-p \bar\theta, && Q_{\frac 12}=x \theta, &&  {\bar Q}_{\frac 12}=x \bar\theta, && J=\frac 12 \theta\bar\theta,
\end{align}
where $(x,p)$ and $(\theta,\bar\theta)$, with ${\theta}^{\dagger}=\bar\theta$, form bosonic and fermionic canonical pairs obeying the Poisson brackets
$\{x,p\}=1$, $\{\theta,\bar\theta \}=-{\rm i}$. Computing the brackets between the functions in (\ref{N2f}) and quantising the result, one obtains the structure relations of $su(1,1|1)$ (Hermitian conjugates are omitted)
\begin{align}\label{alg1}
&
[L_m,L_n]=-{\rm i} (m-n) L_{m+n}, && [L_n,Q_r]=-{\rm i} \left(\frac{n}{2}-r \right) Q_{n+r},
\nonumber\\[2pt]
&
\{Q_r,{\bar Q}_s \}=2 L_{r+s}-2 {\rm i} (r-s) \delta_{r+s,0} J, && [J,Q_r]=\frac 12 Q_r.
\end{align}
It is assumed that the bosonic generators $L_n$ and $J$ are Hermitian, while ${\left(Q_r\right)}^{\dagger}={\bar Q}_r$.
As above, $L_n$, with $n=-1,0,1$, form $sl(2,R)$ subalgebra, $Q_r$, with $r=-\frac 12, \frac 12$, involve the supersymmetry operator and its superconformal partner, while $J$ generates $u(1)$ $R$--symmetry subalgebra.  $(L_n,Q_r,{\bar Q}_r,J)$ have conformal weights $(1,\frac 12,\frac 12,0)$, respectively.

\vspace{0.5cm}

\noindent
{\it 3.2. Acceleration generators versus $(1,2,1)$ supermultiplet}\\

\noindent
Continuing to draw a parallel between the acceleration generators entering an $\mathcal{N}$--extended $\ell$--conformal Galilei superalgebra and irreducible supermultiplets of the $d=1$, $\mathcal{N}$--extended superconformal group, in this subsection we construct an $\mathcal{N}=2$ $\ell$--conformal Galilei superalgebra inspired by a $(1,2,1)$ supermultiplet. An analogy with a real bosonic superfield in $\mathcal{R}^{1|2}$ superspace
\be
\Phi(x,\theta,\bar\theta)=U(x)+ {\rm i} \theta S(x)+{\rm i} \bar\theta {\bar S}(x)+A(x)
\nonumber
\ee
suggests introducing the chain of acceleration generators $({\bf U},{\bf S},\bar{\bf S},{\bf A})$. The bosons  $({\bf U},{\bf A})$ are assumed to be Hermitian operators, while the fermions are Hermitian conjugates of each other, ${\bf S}^{\dagger}=\bar{\bf S}$.

Similarly to the $\mathcal{N}=1$ case, one can either assign the ascending sequence of conformal weights $\left(\ell,\ell+\frac 12,\ell+\frac 12,\ell+1\right)$ to $({\bf U},{\bf S},\bar{\bf S},{\bf A})$, or alternatively choose the descending chain $\left(\ell,\ell-\frac 12,\ell-\frac 12,\ell-1\right)$. The former option implies
\begin{align}\label{n2al}
&
[L_n,{\bf U}_m]=-{\rm i} \left( \ell n-m\right) {\bf U}_{n+m}, && [L_n,{\bf S}_p]=-{\rm i} \left( \left( \ell+\frac 12 \right)n-p \right) {\bf S}_{n+p},
\nonumber\\[2pt]
&
[L_n,{\bar {\bf S}}_p]=-{\rm i} \left( \left( \ell+\frac 12 \right)n-p \right) {\bar{\bf S}}_{n+p}, && [L_n,{\bf A}_q]=-{\rm i} \left( \left(\ell+1 \right)n-q \right) {\bf A}_{n+q},
\end{align}
where the range of values of the indices carried by the generators is prescribed by
\be
{\bf U}_m : m= -\ell,\dots,\ell, \qquad {\bf S}_p : p= -\ell-\frac{1}{2},\dots,\ell+\frac{1}{2}, \qquad {\bf A}_q : q= -\ell-1,\dots,\ell+1.
\nonumber
\ee

In order to specify (anti)commutation relations between $({\bf U},{\bf S},\bar{\bf S},{\bf A})$ and $(Q_r,{\bar Q}_r,J)$, one builds the most general expressions compatible with the conformal weights, Grassmann parities, and index ranges chosen, and analyses the super Jacobi identities. After a straightforward computation one gets
\begin{align}\label{n2al1}
&
[Q_r,{\bf U}_m]={\rm i} {\bf S}_{r+m}, && [J,{\bf U}_m]={\rm i} a(\ell+1) {\bf U}_m,
\nonumber\\[10pt]
&
\{Q_r,{\bar{\bf S}}_p\}=-(1+{\rm i}a) \left((2\ell+1)r-p\right) {\bf U}_{r+p}+{\rm i} {\bf A}_{r+p}, && [J,{\bf A}_q]={\rm i} a(\ell+1) {\bf A}_q,
\nonumber\\[2pt]
&
[Q_r,{\bf A}_q]=(1+{\rm i}a) \left(2(\ell+1)r-q\right) {\bf S}_{r+q}, && [J,{\bf S}_p]=\frac 12 \left(1 +2{\rm i} a (\ell+1)\right) {\bf S}_p,
\end{align}
along with their Hermitian conjugates. Here $a$ is an arbitrary real parameter. Note that for an arbitrary value of $a$ both the bosonic and fermionic acceleration generators do not commute with
$J$. This is to be contrasted with the $su(1,1|1)$ subalgebra, in which only the fermions are sensitive to the $R$--symmetry transformation. It is assumed that $({\bf U},{\bf S},\bar{\bf S},{\bf A})$ (anti)commute with each other. To the best of our knowledge, this $\mathcal{N}=2$ $\ell$--conformal Galilei superalgebra is new.

The descending sequence of conformal weights $\left(\ell,\ell-\frac 12,\ell-\frac 12,\ell-1\right)$ can be considered likewise and the result reads (Hermitian conjugates are omitted)
\begin{align}\label{n2al2}
&
[L_n,{\bf U}_m]=-{\rm i} \left( \ell n-m\right) {\bf U}_{n+m}, && [L_n,{\bf S}_p]=-{\rm i} \left( \left( \ell-\frac 12 \right)n-p \right) {\bf S}_{n+p},
\nonumber\\[2pt]
&
[L_n,{\bf A}_q]=-{\rm i} \left( \left(\ell-1 \right)n-q \right) {\bf A}_{n+q}, && [Q_r,{\bf U}_m]={\rm i}(2\ell r-m) {\bf S}_{r+m}
\nonumber\\[10pt]
&
\{Q_r,{\bar{\bf S}}_p\}=-(1+{\rm i}a){\bf U}_{r+p}+{\rm i}\left((2\ell-1)r-p\right)  {\bf A}_{r+p}, && [Q_r,{\bf A}_q]=(1+{\rm i}a) {\bf S}_{r+q},
\nonumber\\[4pt]
&
[J,{\bf U}_m]=-{\rm i} a \ell {\bf U}_m, && [J,{\bf S}_p]=\frac 12 \left(1 -2{\rm i} a \ell \right) {\bf S}_p,
\nonumber\\[4pt]
&
[J,{\bf A}_q]=-{\rm i} a \ell {\bf A}_q,
&&
\end{align}
where ${\bf U}_m : m= -\ell,\dots,\ell, ~{\bf S}_p : p= -\ell+\frac{1}{2},\dots,\ell-\frac{1}{2}, ~{\bf A}_q : q= -\ell+1,\dots,\ell-1$. In contrast to the previous case, the superalgebra (\ref{n2al2}) is defined for $\ell\geq 1$. Note that at $a=0$ one recovers the $\mathcal{N}=2$ superalgebra introduced in \cite{IM}.

At first glance the variants in (\ref{n2al}), (\ref{n2al1}) and (\ref{n2al2}) are not related.  Given a value of $\ell$, the latter contains two less fermionic operators ${\bf S}_p$ and four less bosonic elements ${\bf A}_q$.  Yet, the replacement $\ell\to\ell+1$ in the descending chain of conformal weights $\left(\ell,\ell-\frac 12,\ell-\frac 12,\ell-1\right)$ yields the ascending one read in the reverse order $\left(\ell+1,\ell+\frac 12,\ell+\frac 12,\ell\right)$. To put it in other words, the $\mathcal{N}=2$ $\ell$--conformal Galilei superalgebra defined by Eqs. (\ref{n2al}), (\ref{n2al1}) contains the same number of generators as the $\mathcal{N}=2$ $(\ell+1)$--conformal Galilei superalgebra in (\ref{n2al2}). The following change
\begin{align}\label{iso}
&
\ell \to\ell+1, && a \to -a,
\nonumber\\[4pt]
&
{\bf U}_{m} \to \frac{{\bf A}_{m}}{1+a^{2}}, &&
{\bf A}_{q} \to - {\bf U}_{q},
\nonumber\\[4pt]
&
{\bf S}_{p}  \to -\frac{\rm i}{1-{\rm i}a}{\bf S}_{p}, && \bar{\bf S}_{p}  \to \frac{\rm i}{1+{\rm i} a}\bar{\bf S}_{p},
\end{align}
establishes the isomorphism of the twin copies.

\vspace{0.5cm}

\noindent
{\it 3.3. Acceleration generators versus $(2,2,0)$ supermultiplet}\\

\noindent
Our next example is an $\mathcal{N}=2$ $\ell$--conformal Galilei superalgebra which links to a $(2,2,0)$ supermultiplet of $SU(1,1|1)$. The first number in the braces implies that one has to consider two real bosonic acceleration generators of the conformal weight $\ell$, or equivalently one {\it complex} operator
\be
[L_n,{\bf U}_m]=-{\rm i} \left( \ell n-m\right) {\bf U}_{n+m}.
\ee
The Hermitian conjugate ${\bar{\bf U}}_m={\left({\bf U}_m \right)}^{\dagger}$, with $m=-\ell,\dots,\ell$, obeys the analogous equation.  The commutator of $Q_r$ from $su(1,1|1)$ and ${\bf U}_m$ yields a complex fermionic generator, which is to be regarded as the analogue of two propagating real fermions entering a $(2,2,0)$ supermultiplet. Assigning the superpartner of ${\bf U}_m$ with the conformal weight $\ell+\frac 12$, one has (Hermitian conjugates are omitted)
\begin{align}\label{QUN2}
&
[Q_r,{\bf U}_m]={\rm i} {\bf S}_{r+m}, && [L_n,{\bf S}_p]=-{\rm i} \left( \left( \ell+\frac 12 \right)n-p \right) {\bf S}_{n+p},
\end{align}
with ${\bf S}_p : p= -\ell-\frac{1}{2},\dots,\ell+\frac{1}{2}$.

As to the remaining (anti)commutators, one considers the most general expressions compatible with the conformal weights, Grassmann
parities, and index ranges chosen, and requires the fulfilment of the super Jacobi identities. The result reads (Hermitian conjugates are omitted)
\bea
&&
\{Q_r,{\bar{\bf S}}_p\}=-2\left((2\ell+1)r-p\right) {\bar{\bf U}}_{r+p}, \qquad [J,{\bf U}_m]=-(\ell+1) {\bf U}_m,
\nonumber\\[2pt]
&&
[J,{\bf S}_p]=-\left(\ell+\frac 12 \right) {\bf S}_p.
\eea
It is assumed that $({\bf U},{\bar{\bf U}},{\bf S},\bar{\bf S})$ (anti)commute with each other.
To the best of our knowledge, this superalgebra has not yet been presented in the literature. Its existence was envisaged by S. Krivonos in \cite{GK} (see a footnote on p. 2).

Turning to the alternative in which the superpartner of the bosonic acceleration generator has the conformal weight $\ell-\frac 12$, one finds
\be
[Q_r,{\bf U}_m]={\rm i}(2\ell r-m) {\bf S}_{r+m}, \qquad [L_n,{\bf S}_p]=-{\rm i} \left( \left( \ell-\frac 12 \right)n-p\right) {\bf S}_{n+p},
\ee
with ${\bf S}_p : p=-\ell+\frac{1}{2},\dots,\ell-\frac{1}{2}$. A similar consideration of other (anti)commutation relations gives
\bea
&&
\{Q_r,{\bar{\bf S}}_p\}=-2 {\bar{\bf U}}_{r+p}, \qquad [J,{\bf U}_m]= \ell {\bf U}_m, \qquad [J,{\bf S}_p]=\left(\ell+\frac 12 \right) {\bf S}_p,
\eea
along with their Hermitian conjugates. Written in different notations, this variant of the $\mathcal{N}=2$ $\ell$--conformal Galilei superalgebra was proposed in \cite{AKT}.
Note that the superalgebra can be derived from the variant above by applying a formal substitution (which similarly to that in Sect. 2 alters the Grassmann parity)
\bea\label{changeN2}
&&
\ell\rightarrow \ell+\frac 12, \quad {\bf U}_{m}\rightarrow {\rm i}\bar{\bf S}_{m}, \quad \bar{\bf U}_{m} \rightarrow {\rm i}{\bf S}_{m}, \quad {\bf S}_{p} \rightarrow -2\bar{\bf U}_{p}, \quad \bar{\bf S}_{p} \rightarrow -2{\bf U}_{p},
\eea
and interchanging commutators with anticommutators where appropriate.

For reader's convenience, realisations of the $\mathcal{N}=2$ $\ell$--conformal Galilei superalgebras in terms of differential operators in superspace are given in Appendix.

Concluding this section, it is worth mentioning that, in contrast to $SU(1,1|1)$ superconformal mechanics, the algebraic considerations above are insensitive to whether $(2,2,0)$ or $(0,2,2)$ supermultiplet is chosen.

\vspace{0.5cm}

\noindent
{\bf 4. $\mathcal{N}=3$ $\ell$--conformal Galilei superalgebras}\\

\noindent
{\it 4.1. Structure relations of $osp(3|2)$}\\

\noindent
$\mathcal{N}=3$ $\ell$--conformal Galilei superalgebras build upon $osp(3|2)$.
The simplest way to establish structure relations of the latter is to consider the Hamiltonian formulation of a free superparticle moving in $\mathcal{R}^{1|3}$ superspace which is parametrized by a real bosonic coordinate $x$ and real fermions $\theta^a$, $a=1,2,3$. Introducing the momentum $p$ canonically conjugate to $x$, regarding $\theta^a$ as self--conjugate variables, and imposing the brackets $\{x,p\}=1$, $\{\theta^a,\theta^b \}=-{\rm i} \delta^{ab}$,
one can verify that the set of functions
\begin{align}
&
L_{-1}=\frac{1}{2} p^2, && L_0=-\frac 12 xp, && L_1=\frac 12 x^2,
\nonumber\\[2pt]
&
Q^a_{-\frac 12}=-p \theta^a, && Q^a_{\frac 12}=x \theta^a, && J^a=-\frac{{\rm i}}{2} \epsilon^{abc} \theta^b \theta^c,
\end{align}
where $\epsilon^{abc}$ is the Levi--Civita symbol, forms a closed superalgebra. Promoting the brackets to (anti)commutators in the usual way, one obtains the structure relations of $osp(3|2)$
\begin{align}\label{alg1}
&
[L_m,L_n]=-{\rm i} (m-n) L_{m+n}, && [L_n,Q^a_r]=-{\rm i} \left(\frac{n}{2}-r \right) Q^a_{n+r},
\nonumber\\[4pt]
&
\{Q^a_r,Q^b_s \}=2 L_{r+s} \delta^{ab}+(r-s) \delta_{r+s,0} \epsilon^{abc} J^c,  && [J^a,Q^b_r]={\rm i} \epsilon^{abc} Q^c_r,
\nonumber\\[4pt]
&
[J^a,J^b]= {\rm i} \epsilon^{abc} J^c. &&
\end{align}
As above, $L_n$, with $n=-1,0,1$, generate the conformal subalgebra $sl(2,R)$, $Q^a_r$, with $r=-\frac 12, \frac 12$ and $a=1,2,3$, encode the supersymmetry operator and its superconformal partner, while $J^a$, with $a=1,2,3$, generate $so(3)$ $R$--symmetry subalgebra. As follows from (\ref{alg1}), $(L_n,Q^a_r,J^a)$ have conformal weights $(1,\frac 12,0)$, respectively. All operators are assumed to be Hermitian.

Similarly to the analysis above, one can try to construct $\mathcal{N}=3$ $\ell$--conformal Galilei superalgebras which link to $(1,3,2)$ and $(3,3,0)$ supermultiplets (the options $(2,3,1)$ and $(0,3,3)$ follow by redefinition of $\ell$). In the former case, the starting point would be the triplet of acceleration generators $({\bf U},{\bf S}^a,{\bf A})$, where ${\bf U}$ and its superpartner ${\bf S}^a$, with $a=1,2,3$, are real, while ${\bf A}$ is complex. Assigning them with the conformal weights $(\ell,\ell+\frac 12,\ell+1)$ and analysing the most general structure relations compatible with the Grassmann parities, one finds out that the super Jacobi identities do not hold in the sector involving two operators of the type $Q^a_r$ and one acceleration generator from the triplet $({\bf U},{\bf S}^a,{\bf A})$. Similar problem occurs for a $(3,3,0)$ supermultiplet, which would be represented by the Hermitian operators $({\bf U}^a,{\bf S}^a)$, with $a=1,2,3$.

A natural way out is to turn to reducible supermultiplets of $OSp(3|2)$, the simplest of which is a real bosonic scalar superfield defined in $\mathcal{R}^{1|3}$ superspace
\be
\Phi(x,\theta)=U(x)+ \theta^a S^a (x)+\frac 12 \theta^a \theta^b \epsilon^{abc} A^c (x)+\frac{1}{3!} \epsilon^{abc} \theta^a \theta^b \theta^c P(x)
\nonumber
\ee
where $U(x)$, $A^c (x)$ are Grassmann--even and $S^a (x)$, $P(x)$ are Grassmann--odd. Below we construct two new $\mathcal{N}=3$ $\ell$--conformal Galilei superalgebras, acceleration generators of which fit the reducible supermultiplet $(U,S^a,A^a,P)$.

\vspace{0.5cm}

\noindent
{\it 4.2. Acceleration generators of conformal weights $(\ell,\ell+\frac 12,\ell+1,\ell+\frac 32)$}\\

\noindent
As usual, one starts with a Hermitian operator ${\bf U}_m$ of the conformal weight $\ell$
\be
[L_n,{\bf U}_m]=-{\rm i} \left( \ell n-m\right) {\bf U}_{n+m}.
\ee
As ${\bf U}_m$ does not carry $so(3)$ indices, it commutes with $J^a$. The commutator of ${\bf U}_m$ with the fermionic operators $Q^a_r$ is Grassmann--odd and, hence, should be regarded as a new fermionic Hermitian generator of the conformal weight $\ell+\frac 12$
\be\label{QU}
[Q^a_r,{\bf U}_m]={\rm i} {\bf S}^a_{r+m}, \quad [L_n,{\bf S}^a_p]=-{\rm i} \left( \left( \ell+\frac 12 \right)n-p \right) {\bf S}^a_{n+p}, \quad [J^a,{\bf S}^b_p]={\rm i} \epsilon^{abc} {\bf S}^c_p,
\ee
with ${\bf S}^a_p : p= -\ell-\frac{1}{2},\dots,\ell+\frac{1}{2}$.

The structure of $so(3)$ indices entering the anticommutator $\{Q^a_r,{\bf S}^b_s \}$ and the fact that it is Grassmann--even suggest the structure relation
\be\label{QS}
\{Q^a_r,{\bf S}^b_p \}=-\left( \left( 2\ell+1 \right)r-p \right) \delta^{ab} {\bf U}_{r+p}+\epsilon^{abc} {\bf A}^c_{r+p},
\ee
where ${\bf A}^a_q : q=-\ell-1,\dots,\ell+1$, with $a=1,2,3$, is a new bosonic Hermitian generator to be included into the ensuing superalgebra. The factor $-\left( \left( 2\ell+1 \right)r-p \right)$ entering the first term is so designed as to keep the index carried by ${\bf U}$ from leaving the range $-\ell,\dots,\ell$, as well as to guarantee the fulfilment of the super Jacobi identities.

In its turn, the extra bosonic generator ${\bf A}^a_q$ has the conformal weight $\ell+1$ and carries $so(3)$ vector index, which imply
\be
[L_n, {\bf A}^a_q]=-{\rm i} \left( \left(\ell+1 \right)n-q\right) {\bf A}^a_{n+q}, \qquad [J^a,{\bf A}^b_q]={\rm i} \epsilon^{abc} {\bf A}^c_q,
\ee
while the commutator of $Q^a_r$ and ${\bf A}^b_q$ produces a new fermionic Hermitian generator ${\bf P}_{s}$
\be\label{QA}
[Q^a_r, {\bf A}^b_q]={\rm i} \delta^{ab} {\bf P}_{r+q}+{\rm i} \left( \left(2\ell+2 \right)r-q\right) \epsilon^{abc} {\bf S}^c_{r+q},
\ee
with ${\bf P}_{s} : s=-\ell-\frac 32,\dots,\ell+\frac 32$. Similarly to (\ref{QS}), the meaning of the factor ${\rm i}\left( \left(2\ell+2 \right)r-q\right)$ entering the second term is to
balance in a proper way the range of values of indices carried by the acceleration generators on both sides of the equality (\ref{QA}) as well as to ensure the super Jacobi identities.

Finally, the conformal weights of $Q^a_r$ and ${\bf A}^b_m$ add up to yield that of ${\bf P}_{s}$
\be
[L_n,{\bf P}_s]=-{\rm i}  \left( \left( \ell+\frac 32 \right)n-s \right) {\bf P}_{n+s},
\ee
while the structure of indices entering the anticommutator $\{Q^a_r, {\bf P}_s\}$ implies
\be\label{QP}
\{Q^a_r, {\bf P}_s\}=-\left( \left(2\ell+3 \right)r-s\right) {\bf A}^a_{r+s}.
\ee
The factor $-\left( \left(2\ell+3 \right)r-s\right)$ harmonises the range of values of indices on both sides of (\ref{QP}) and conforms to super Jacobi identities.
It is assumed that $[J^a,{\bf P}_{r}]=0$ and the acceleration generators commute with each other.

\vspace{0.5cm}

\noindent
{\it 4.3. Acceleration generators of conformal weights $(\ell,\ell-\frac 12,\ell-1,\ell-\frac 32)$}\\

\noindent
A twin copy of the $\mathcal{N}=3$ $\ell$--conformal Galilei superalgebra in the preceding section arises if one assigns $({\bf U},{\bf S}^a,{\bf A}^a,{\bf P})$ with the conformal weights $(\ell,\ell-\frac 12,\ell-1,\ell-\frac 32)$. Omitting the details, we display below the corresponding structure relations (vanishing (anti)commutators are omitted)
\begin{align}
&
[L_n,{\bf U}_m]=-{\rm i} \left( \ell n-m\right) {\bf U}_{n+m}, && [Q^a_r,{\bf U}_m]={\rm i} (2 \ell r-m) {\bf S}^a_{r+m},
\nonumber\\[2pt]
&
[L_n,{\bf S}^a_p]=-{\rm i} \left( \left( \ell-\frac 12 \right)n-p \right) {\bf S}^a_{n+p}, && [J^a,{\bf S}^b_p]={\rm i} \epsilon^{abc} {\bf S}^c_p,
\nonumber\\[2pt]
&
\{Q^a_r,{\bf S}^b_p \}=-\delta^{ab} {\bf U}_{r+p}+\left( \left( 2\ell-1 \right)r-p \right)\epsilon^{abc} {\bf A}^c_{r+p}, &&
[J^a,{\bf A}^b_q]={\rm i} \epsilon^{abc} {\bf A}^c_q,
\nonumber\\[2pt]
&
[L_n, {\bf A}^a_q]=-{\rm i} \left( \left(\ell-1 \right)n-q\right) {\bf A}^a_{n+q}, && [L_n,{\bf P}_s]=-{\rm i}  \left( \left( \ell-\frac 32 \right)n-s \right) {\bf P}_{n+s},
\nonumber\\[2pt]
&
[Q^a_r, {\bf A}^b_q]={\rm i} \left( \left(2\ell-2 \right)r-q\right) \delta^{ab} {\bf P}_{r+q}+{\rm i}  \epsilon^{abc} {\bf S}^c_{r+q}, && \{Q^a_r, {\bf P}_s\}=-{\bf A}^a_{r+s},
\end{align}
where it is assumed that $\ell \geq \frac 32$. The range of values of the lower indices carried by the acceleration generators is specified by
\begin{align}
&
{\bf U}_m: m=-\ell,\ldots,\ell, &&  {\bf S}^a_p: p= -\ell+\frac{1}{2},\ldots,\ell-\frac{1}{2},
\nonumber\\[2pt]
&
{\bf A}^a_q: q = -\ell+1, \ldots, \ell-1, && {\bf P}_s: s=-\ell+\frac{3}{2},\ldots,\ell-\frac{3}{2}.
\nonumber
\end{align}
It is straightforward to verify that the super Jacobi identities are satisfied. Note that the two variants of the $\mathcal{N}=3$ $l$--conformal Galilei superalgebra obtained in this section are related to each other by a duality transformation similar to \eqref{changeN1} and \eqref{changeN2}
\bea\label{changeN3}
&&
\ell \to \ell+\frac 32, \quad {\bf U}_{m} \to {\rm i} {\bf P}_{m}, \quad {\bf P}_{s} \to {\bf U}_{s}, \quad {\bf S}_{p}^{a} \to -{\bf A}_{p}^{a}, \quad {\bf A}_{q}^{a} \to -\mathrm{i}{\bf S}_{q}^{a}.
\eea

Realisations of the $\mathcal{N}=3$ $\ell$--conformal Galilei superalgebras in terms of differential operators in superspace can be found in Appendix.

\vspace{0.5cm}

\noindent
{\bf 5. Conclusion}\\

\noindent
To summarise, in this work the issue of constructing $\mathcal{N}=1,2,3$ supersymmetric extensions of the $\ell$--conformal Galilei algebra was reconsidered along the lines in \cite{GK}. A new $\mathcal{N}=1$ version, two new $\mathcal{N}=2$ superalgebras, and two new $\mathcal{N}=3$ variants were built. Their realisations in terms of differential operators in superspace were given.

Turning to possible further developments, it would be interesting to study dynamical realisations of the superalgebras proposed above. Only a limited number of such examples are available (see e.g. \cite{AG2,GL,AG} for $\ell=\frac 12$ and \cite{IM1,IM2} for an arbitrary value of $\ell$) and a deeper understanding of peculiarities of an $\mathcal{N}$--extended $\ell$--conformal Galilei supersymmetry is desirable. In particular, it would be interesting to build an $\mathcal{N}=2$ model that would assign a clear physical meaning to the parameter $a$ introduced in Sect. 3.

Above it was assumed that the acceleration generators commute with each other. Analysis of possible central extensions is an interesting task. A related open problem is the construction of Casimir elements.

In this work, only finite--dimensional superalgebras were considered. Allowing the index $n$ of $L_n$ to carry any integer value, the index $r$ characterising $Q_r$ to take any half--integer value and similarly for the acceleration generators, one would automatically generate infinite--dimensional extensions. A possibility to use them within the context of superconformal field theory is worth studying.

As is well known, the Galilei algebra can be obtained from the Poincar\'e algebra by applying the non-–relativistic contraction. A similar derivation of the $\ell$--conformal Galilei algebra is available only for $\ell=\frac 12$ \cite{Barut} and $\ell=1$ \cite{LSZ}, which relies upon $so(2,d+1)$. It would be interesting to study in which way the full tower of acceleration generators for an arbitrary value of $\ell$ can be obtained from $so(p,q)$ (with properly adjusted $p$ and $q$) by applying the In\"on\"u–-Wigner contraction and
whether such a consideration can be extended to encompass the supersymmetric cases studied in this work. 

Finally, it would be interesting to extend the analysis in this work to $su(1,1|n)$, $osp(n|2)$, and $osp(4^{*}|2n)$ superconformal algebras.

\vspace{0.5cm}

\noindent{\bf Acknowledgements}\\

\noindent
This work is supported by the Russian Science Foundation, grant No 19-11-00005.

\vspace{0.5cm}

\noindent
{\bf Appendix: Realisations in superspace }

\vskip 0.5cm
\noindent
Let us first discuss the $\mathcal{N}=1$ case.
As was demonstrated in \cite{MM}, the differential operators
\bea\label{realN=1}
\begin{aligned}
&
Q_{-\frac 12} = \frac{\partial}{\partial\theta} + {\rm i}\theta\frac{\partial}{\partial t}, && Q_{\frac 12} = t\frac{\partial}{\partial\theta} + {\rm i} t \theta\frac{\partial}{\partial t} + 2 {\rm i} \ell \theta x_{\alpha} \frac{\partial}{\partial x_{\alpha}},
\\[2pt]
&
L_{-1} = {\rm i}\frac{\partial}{\partial t}, && L_{0} = {\rm i}t\frac{\partial}{\partial t} + \frac{{\rm i}}{2}\theta\frac{\partial}{\partial\theta} + {\rm i}\ell x_{\alpha}\frac{\partial}{\partial x_{\alpha}},
\\[2pt]
&
L_{1} = {\rm i} t^{2}\frac{\partial}{\partial t} + {\rm i} t \theta\frac{\partial}{\partial\theta} + 2{\rm i}\ell t x_{\alpha}\frac{\partial}{\partial x_{\alpha}}, && U_{m}^{\alpha}={\rm i}t^{m+\ell}\frac{\partial}{\partial x_{\alpha}},\quad S_{p}^{\alpha} = -{\rm i}\theta t^{p+\ell-\frac 12}\frac{\partial}{\partial x_{\alpha}},
\nonumber
\end{aligned}
\eea
form a representation of the $\mathcal{N}=1$ $\ell$--conformal Galilei superalgebra for the case when the generators ${\bf U}_{m}$ and ${\bf S}_{p}$ have the conformal weights $\ell$ and $\ell-\frac 12$, respectively. Here $(t,x_{\alpha})$, with $\alpha=1,\dots,d$, are real bosonic coordinates and $\theta$ is a real fermion. A realisation of the twin copy algebra, which involves the acceleration generators of conformal weights $(\ell,\ell+\frac 12)$, is achieved by considering the inverse of the transformation \eqref{changeN1} and changing $x_{\alpha}$ by a new fermionic variable $\psi_{\alpha}$.

Proceeding to the $\mathcal{N}=2$ case, the superalgebra associated with a supermultiplet $(1,2,1)$ and the descending sequence of conformal weights can be realised as follows
\bea
&&
\begin{aligned}
&
Q_{-\frac 12} = \frac{\partial}{\partial\theta} + \mathrm{i}\bar{\theta}\frac{\partial}{\partial t}, && Q_{\frac 12} = t\frac{\partial}{\partial\theta} + \mathrm{i}t\bar{\theta}\frac{\partial}{\partial t} + 2\mathrm{i}\ell(1 - \mathrm{i} a) \bar{\theta} x_{\alpha}\frac{\partial}{\partial x_{\alpha}} + \mathrm{i}\bar{\theta}\theta \frac{\partial}{\partial\theta},
\\[2pt]
&
\bar{Q}_{- \frac 12} = \frac{\partial}{\partial\bar{\theta}} + \mathrm{i}\theta\frac{\partial}{\partial t}, &&
\bar{Q}_{\frac 12} = t\frac{\partial}{\partial\bar{\theta}} + \mathrm{i}t\theta\frac{\partial}{\partial t} + 2\mathrm{i}\ell(1 + \mathrm{i} a) \theta x_{\alpha}\frac{\partial}{\partial x_{\alpha}} + \mathrm{i}\theta\bar{\theta} \frac{\partial}{\partial\bar{\theta}},
\\[2pt]
&
L_{-1} = \mathrm{i}\frac{\partial}{\partial t}, && L_{0} = \mathrm{i}t\frac{\partial}{\partial t} + \mathrm{i} \ell x_{\alpha}\frac{\partial}{\partial x_{\alpha}} + \frac{\mathrm{i}}{2}\theta\frac{\partial}{\partial\theta} + \frac{\mathrm{i}}{2}\bar{\theta}\frac{\partial}{\partial\bar{\theta}},
\\[2pt]
&
J = \frac{1}{2}\bar{\theta}\frac{\partial}{\partial\bar{\theta}} - \frac{1}{2}\theta\frac{\partial}{\partial\theta} + \mathrm{i}a\ell x_{\alpha}\frac{\partial}{\partial x_{\alpha}}, && L_{1} = \mathrm{i}t^{2}\frac{\partial}{\partial t} + 2\mathrm{i}\ell (t + a\theta\bar{\theta}) x_{\alpha}\frac{\partial}{\partial x_{\alpha}} + \mathrm{i} t \theta\frac{\partial}{\partial\theta} + \mathrm{i} t\bar{\theta}\frac{\partial}{\partial\bar{\theta}},
\\[2pt]
&
U_{m}^{\alpha} = \mathrm{i}(t + a\theta\bar{\theta})^{\ell+m}\frac{\partial}{\partial x_{\alpha}}, && S_{p}^{\alpha} = - {\rm i} (1- {\rm i}a)\bar{\theta} t^{\ell+p-1/2}\frac{\partial}{\partial x_{\alpha}},
\\[2pt]
&
A_{q}^{\alpha} = {\rm i}(1+a^{2})t^{\ell+q-1}\bar{\theta}\theta \frac{\partial}{\partial x_{\alpha}}, && \bar{S}_{p}^{\alpha} = - {\rm i} (1+ {\rm i}a)\theta t^{\ell+p-1/2}\frac{\partial}{\partial x_{\alpha}},
\end{aligned}
\nonumber
\eea
where $(t,x_{\alpha})$, with $\alpha=1,\dots,d$, are real bosonic coordinates, while $(\theta,\bar\theta)$ are complex conjugate fermions. Note that for $a=0$, the representation correctly reduces to that in \cite{IM}. A variant corresponding to the ascending sequence of conformal weights is obtained via the isomorphism (\ref{iso}).

The $\mathcal{N}=2$ superalgebra associated with a $(2,2,0)$ supermultiplet and the descending sequence of conformal weights can be represented by the differential operators
\bea
&&
Q_{-1/2} = \frac{\partial}{\partial\theta} + \mathrm{i}\bar{\theta}\frac{\partial}{\partial t}, \qquad\; Q_{1/2} = t\frac{\partial}{\partial\theta} + \mathrm{i}t\bar{\theta}\frac{\partial}{\partial t} + 4\mathrm{i}\ell \bar{\theta} \bar{z}_{\alpha}\frac{\partial}{\partial\bar{z}_{\alpha}} + \mathrm{i}\bar{\theta}\theta \frac{\partial}{\partial\theta},
\nonumber
\\[2pt]
&&
\bar{Q}_{-1/2} = \frac{\partial}{\partial\bar{\theta}} + \mathrm{i}\theta\frac{\partial}{\partial t}, \qquad\; \bar{Q}_{1/2} = t\frac{\partial}{\partial\bar{\theta}} + \mathrm{i}t\theta\frac{\partial}{\partial t} + 4\mathrm{i}\ell \theta z_{\alpha}\frac{\partial}{\partial z_{\alpha}} + \mathrm{i}\theta\bar{\theta} \frac{\partial}{\partial\bar{\theta}},
\nonumber
\\[2pt]
&&
L_{-1} = \mathrm{i}\frac{\partial}{\partial t}, \qquad\qquad\qquad\; L_{0} = \mathrm{i}t\frac{\partial}{\partial t} + \mathrm{i} \ell z_{\alpha}\frac{\partial}{\partial z_{\alpha}} + \mathrm{i} \ell \bar{z}_{\alpha}\frac{\partial}{\partial \bar{z}_{\alpha}} + \frac{\mathrm{i}}{2}\theta\frac{\partial}{\partial\theta} + \frac{\mathrm{i}}{2}\bar{\theta}\frac{\partial}{\partial\bar{\theta}},
\nonumber
\\[2pt]
&&
L_{1} = \mathrm{i}t^{2}\frac{\partial}{\partial t} + 2\mathrm{i} \ell (t - \mathrm{i}\theta\bar{\theta}) z_{\alpha}\frac{\partial}{\partial z_{\alpha}} + 2\mathrm{i} \ell(t + \mathrm{i}\theta\bar{\theta}) \bar{z}_{\alpha}\frac{\partial}{\partial \bar{z}_{\alpha}} + \mathrm{i} t \theta\frac{\partial}{\partial\theta} + \mathrm{i} t\bar{\theta}\frac{\partial}{\partial\bar{\theta}},
\nonumber
\\[2pt]
&&
U_{m}^{\alpha} =  \mathrm{i}(t + i\theta\bar{\theta})^{\ell+m}\frac{\partial}{\partial \bar{z}_{\alpha}}, \quad J = \frac{1}{2}\bar{\theta}\frac{\partial}{\partial\bar{\theta}} - \frac{1}{2}\theta\frac{\partial}{\partial\theta} + \ell z_{\alpha}\frac{\partial}{\partial z_{\alpha}} - \ell \bar{z}_{\alpha}\frac{\partial}{\partial \bar{z}_{\alpha}},
\nonumber
\\[2pt]
&&
\bar{U}_{m}^{\alpha} = \mathrm{i}(t - i\theta\bar{\theta})^{\ell+m}\frac{\partial}{\partial z_{\alpha}}, \quad S_{p}^{\alpha} = -2\mathrm{i}t^{\ell+p-1/2}\bar{\theta}\frac{\partial}{\partial\bar{z}_{\alpha}}, \;
\bar{S}_{p}^{\alpha} = -2\mathrm{i}t^{\ell+p-1/2}\theta\frac{\partial}{\partial z_{\alpha}}.
\nonumber
\eea
where $t$ is a real bosonic coordinate, $(z_{\alpha},\bar{z}_{\alpha})$, with $\bar{z}_{\alpha} = (z_{\alpha})^{\dagger}$, $\alpha=1,\dots,d$, are complex bosonic variables, and $(\theta,\bar\theta)$ are complex conjugate fermions. In order to describe a twin copy, it suffices to apply \eqref{changeN2} and change $z_{\alpha}$ by a complex fermionic analogue $z_\alpha \to \psi_{\alpha}$.

The $\mathcal{N}=3$ $\ell$--conformal Galilei superalgebra associated with the descending sequence of conformal weights $(\ell,\ell-\frac 12,\ell-1,\ell-\frac 32)$ can be represented by the differential operators
\bea
&&
\begin{aligned}
&
Q_{-\frac{1}{2}}^{a} = \frac{\partial}{\partial\theta^{a}} + {\rm i}\theta^{a}\frac{\partial}{\partial t}, && Q_{\frac{1}{2}}^{a} = tQ_{-\frac{1}{2}}^{a} + {\rm i}\theta^{a}\theta^{b}\frac{\partial}{\partial\theta^{b}} + 2{\rm i}\ell\theta^{a}x_{\alpha}\frac{\partial}{\partial x_{\alpha}},
\\[2pt]
&
L_{-1} = {\rm i}\frac{\partial}{\partial t}, && L_{0} = {\rm i}t\frac{\partial}{\partial t} + {\rm i}\ell x_{\alpha}\frac{\partial}{\partial x_\alpha} + \frac{{\rm i}}{2}\theta^{a}\frac{\partial}{\partial\theta^{a}},
\\[2pt]
&
J^{a} = -{\rm i}\epsilon^{abc}\theta^{b}\frac{\partial}{\partial\theta^{c}}, && L_{1} = {\rm i}t^{2}\frac{\partial}{\partial t} + 2{\rm i}\ell tx_{\alpha}\frac{\partial}{\partial x_\alpha} + {\rm i}t\theta^{a}\frac{\partial}{\partial\theta^{a}},
\\[2pt]
&
U_{m}^{\alpha} = {\rm i} t^{m+\ell}\frac{\partial}{\partial x_{\alpha}}, && A_{m}^{a,\alpha} = -\frac{1}{2}\epsilon^{abc}\theta^{b}\theta^{c}t^{m + \ell - 1}\frac{\partial}{\partial x_\alpha},
\\[2pt]
&
S_{r}^{a,\alpha} = -{\rm i} \theta^{a}t^{r+\ell-\frac{1}{2}}\frac{\partial}{\partial x_{\alpha}}, && P_{r}^{\alpha} = \frac{ 1}{6}\epsilon^{abc}\theta^{a}\theta^{b}\theta^{c}t^{r+\ell-\frac{3}{2}}\frac{\partial}{\partial x_\alpha},
\end{aligned}
\nonumber
\eea
where $(t,x_{\alpha})$, with $\alpha=1,\dots,d$, are real bosonic coordinates and $\theta^{a}$, with $a=1,2,3$, are real fermions. The twin copy, which builds upon the ascending sequence of conformal weights, follows from
\eqref{changeN3} and $x_{\alpha} \to \psi_{\alpha}$, where $\psi_\alpha$ is a fermionic variable.

\end{document}